\newcommand{\ud}{\rm d}
\newcommand{\un}{~\mathrm}

\documentclass[twocolumn,showpacs,preprintnumbers,amsmath,amssymb]{revtex4}
\usepackage{graphicx}% Include figure files
\usepackage{dcolumn}% Align table columns on decimal point
\usepackage{bm}% bold math
\usepackage{amsmath}

\begin{document}

\title{Multi-scale clustering in granular surface flows}
\author{D. Bonamy}
\author{F. Daviaud}
\author{L. Laurent}
\author{M. Bonetti}
\author{J.P. Bouchaud}
\affiliation{Service de Physique de l'Etat Condens\'e, CEA Saclay, 91191 Gif
sur Yvette, France}

%\date{\today}
%\vspace{20mm}

\begin{abstract}

We investigate steady granular surface flows in a rotating drum and demonstrate 
the existence of rigid clusters of grains embedded in the flowing layer. 
These clusters are fractal and their size is power-law distributed from the grain size scale 
up to the thickness of the flowing layer. The implications of the absence of a characteristic 
length scale on available theoretical models of dense granular flows are discussed. Finally, we 
suggest a
possible explanation of the difference between velocity profiles observed in surface flows and in flows down a rough inclined plane.

\end{abstract}

\maketitle
\vspace{1.5cm}

\address{}
\date{\today}
\maketitle

% body of paper here
Granular materials share properties with both usual liquids and solids. They can form an inclined free surface without flowing, but when the angle of the free surface exceeds some threshold value, an avalanche occurs. Global behaviours can be described by models derived from fluid mechanics~\cite{Savage89,Douady99} or non linear physics~\cite{Bouchaud94,Aranson01}. 
However, some experimental results remain far from being understood.
Detailed measurements of the mean density profile and the mean tangential velocity profile 
obtained in  three-dimensional flow (3D) by NMR~\cite{Nagakawa93} and in 
two-dimensional flows (2D) by direct image analysis~\cite{Rajchenbach00,Bonamy01}, show strong evidence that the relation between stress and strain is {\it non local}: 
(i) the velocity gradient is found to be constant in the flowing layer whereas momentum balance predicts a linear variation of the shear stress with depth~\cite{Mills99}; 
(ii)  the velocity gradient presents a different scaling with the depth for dense granular flows down a rough inclined plane~\cite{Drake90,Azanza97,Pouliquen99} indicating the non local influence of boundary conditions on internal rheology inside the flowing layer; 
(iii) the velocity gradient does not vanish at the free surface at variance with typical fluids.
Very recent models have tried to account for some non local effects~\cite{Rajchenbach00,Mills99,Andreotti01}.
We report here the first experimental evidence of rigid clusters of grains embedded in the flow
and characterize their geometrical and statistical properties. 
Although clustering instabilities driven by the inelasticity of grain collision are well known
in granular gases~\cite{Goldhirsch98,Falcon99}, they have never been observed in dense surface flows. We find that these clusters are fractal and their size is power-law distributed from the microscopic scale -- the diameter of a grain -- up to the macroscopic scale -- the flowing layer thickness. Therefore, no characteristic correlation length can be defined in the flowing layer. 
Almost $50\%$ of flowing beads belong to these objects. We discuss recently proposed non local models in the light of these experimental observations, and propose a argument to explain the differences observed between granular surface flows and granular flow down a rough inclined plane.

\begin{figure}
\centering
\hspace{0.15\columnwidth}\includegraphics[ width=0.98\columnwidth]{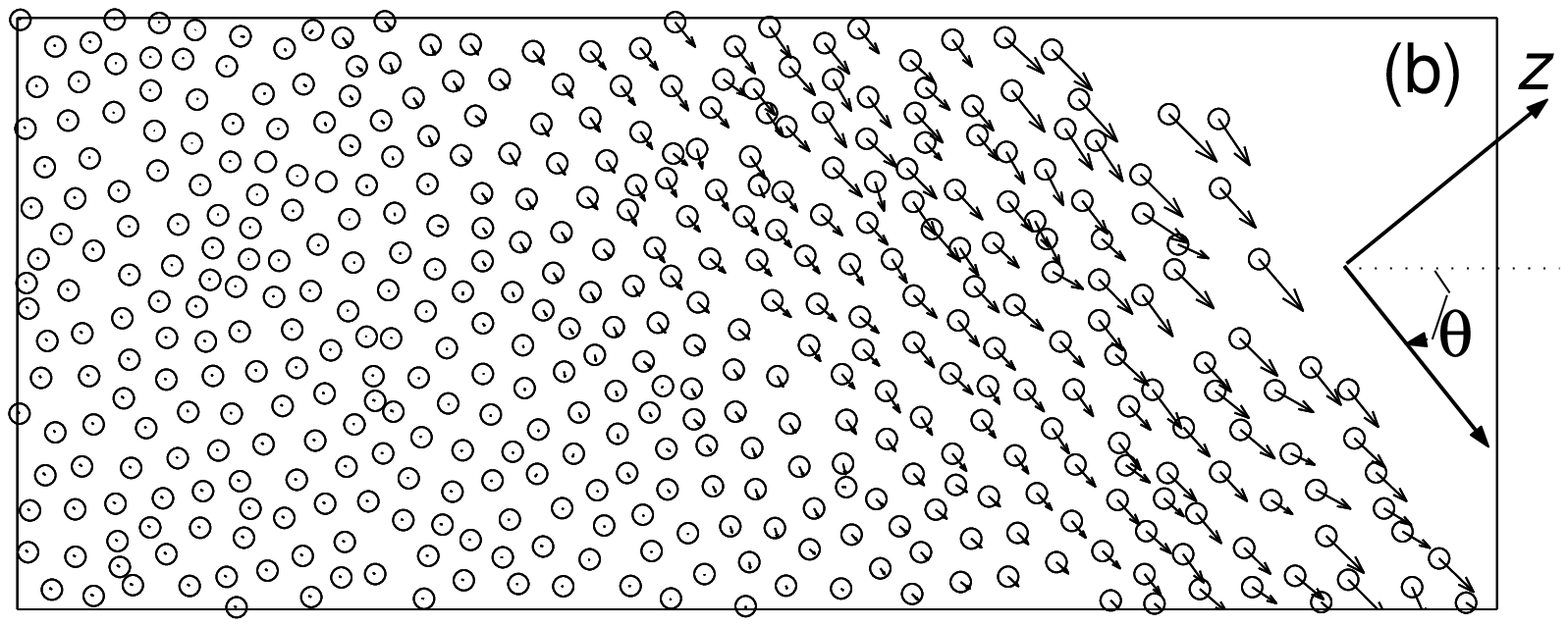}
\includegraphics[ width=0.49\columnwidth]{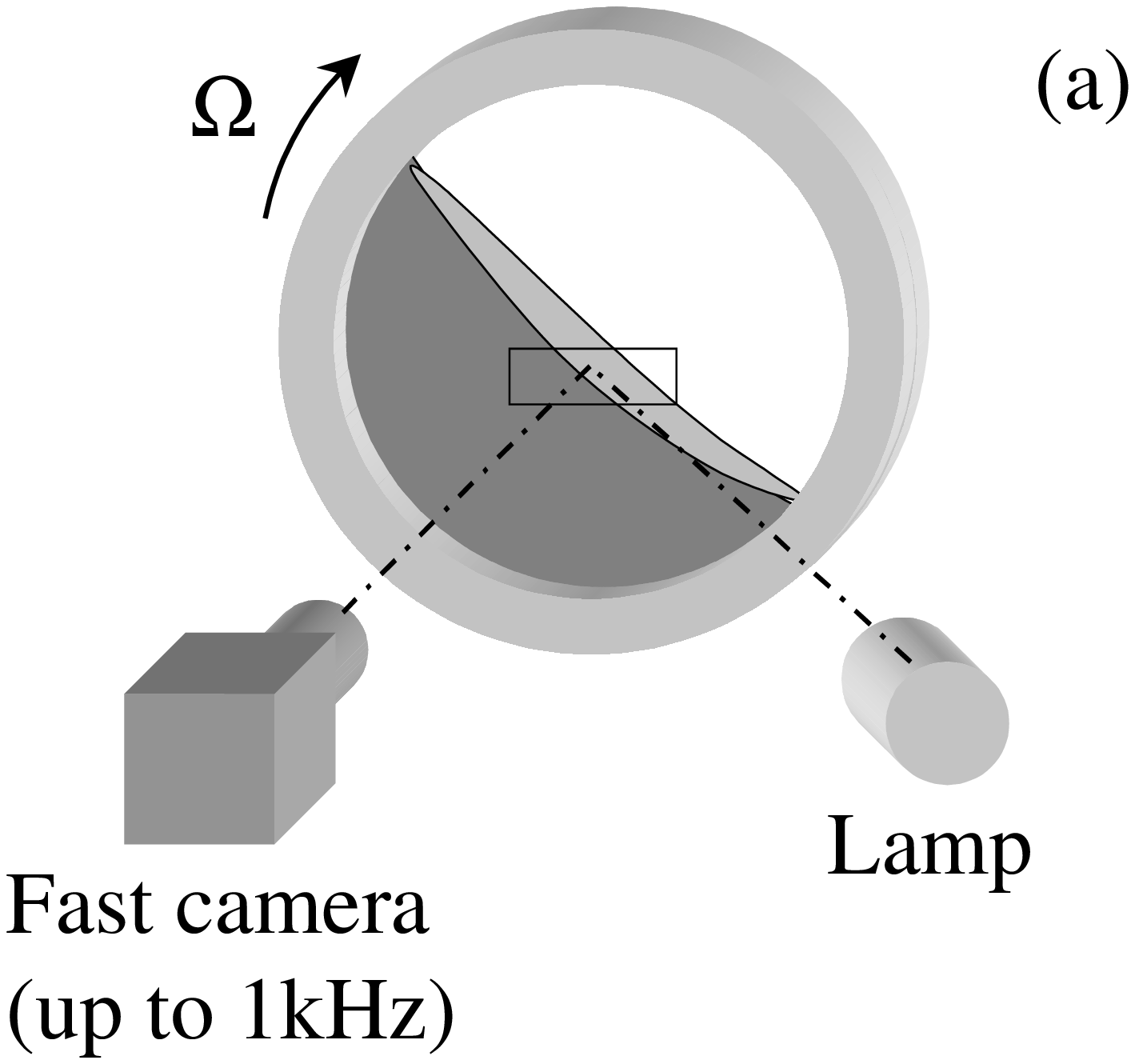}
\includegraphics[ width=0.49\columnwidth]{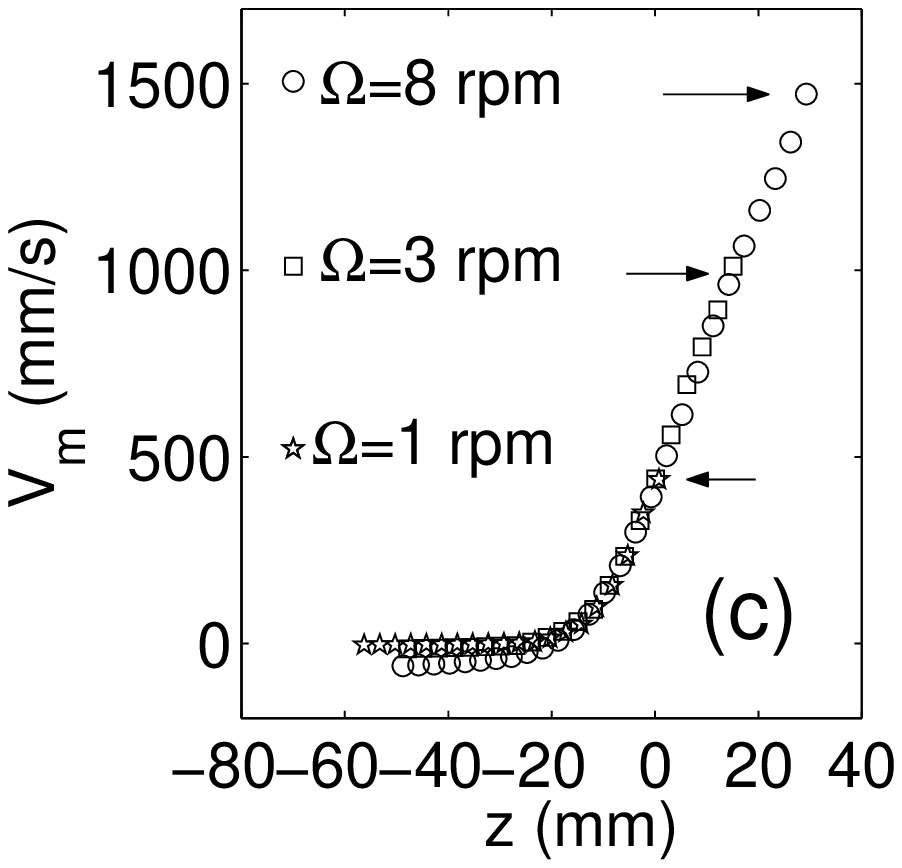}
\caption{(a) Sketch of the experimental setup. (b) A typical instanteous velocity field obtained for quasi-2D flows (geometry A) with $\Omega=8\un{rpm}$ in the rectangular box of (a). (c) Velocity profile measured at the center of the drum for 3 different rotation speed: $\Omega=1\un{rpm}$, $\Omega=3\un{rpm}$ and $\Omega=8\un{rpm}$. The arrows indicates the end of each profile. The precision on each point is better than the point size. Velocity profiles are linear in the flowing layer with a gradient independent of $\Omega$, {\em i.e.} independent of both the local slope $\theta$ and the flowing layer thickness $R$.} 
\label{fig1}
\end{figure}

{\em Experimental setup:} The experimental setup is illustrated in Fig.~\ref{fig1}a. It consists
in a duralumin rotating drum of diameter $D_0=45\un{cm}$ and variable gap, half-filled with steel beads of diameter $d=3\pm0.05\un{mm}$. The use of weakly dissipative beads of millimetric size prevent any clustering due to material-specific effects such as capillary or electrostatic effects.  
Two drum thickness are used:
\begin{itemize}
\item[(A)] a gap of $7\un{mm}$ so that a quasi-2D packing is obtained but with a local 3D microscopic disordered structure. The fast camera allows then to track the beads actually seen through the transparent side wall of the tumbler: around $40\%$ of the beads are hidden and cannot be tracked.

\item[(B)] a gap of $3\un{mm}$ so that a pure 2D packing is obtained and all the beads can be tracked. Additional metallic obstacles (bound to the inside of the drum) have been added in the static part of the packing to prevent any 2D ordering effects which would induce non generic 
effects. The main drawback is that the measurements can be only made during the time when the obstacles are in the bottom part of the drum. 
\end{itemize}
\noindent In the recorded region, located at the center of the drum, the flowing layer thickness $R$ and the mean angle $\theta$ of the flow can be changed by varying the rotation speed $\Omega$. Regimes obtained for $\Omega$ varying from $1\un{rpm}$ to $8\un{rpm}$ are investigated. 
On this range of rotating speeds, surface flows are steady and inertial effect
are negligible (the Froude number $Fr=\Omega D_0/2g \leq 0.01$ where $g$ is the gravity constant).

The beads are lighted via a continuous halogen lamp. Sequences of $200$ frames are recorded via a 
fast camera at a sampling rate  of $1\un{kHz}$. The recorded region size in pixel is $480\times234$, one pixel corresponding to $0.227\un{mm}$. Frame processing allows us to obtain the position of the
center of mass of the beads seen through the transparent side wall of the tumbler. Since the image 
of a single bead is made up of about $23\pm3$ pixels (depending on the distance to the side wall),
 the errors on the determination of bead location is about $150\,\mu{\rm m}$~\cite{evalError}.
Tracking each bead on ten successive frames allows us to evaluate its velocity averaged on $10\un{ms}$ (see Fig.~\ref{fig1}b). The mean angle $\theta$ of the flow is calculated from the 
averaging of the velocity of all the beads on the whole sequence of $200$ frames. The images are 
then divided into layers 1 bead diameter wide parallel to the flow. The mean tangential velocity 
$V_m(z)$ of the layer at depth $z$ is then defined as the average of the velocity of all the beads of the sequence of $200$ frames whose center of mass is inside the layer.

For both geometries (A) and (B), the granular surface flow indeed presents a clear linear velocity profile (see Fig.~\ref{fig1}c for the quasi 2D case) with a velocity gradient $\dot{\gamma}={\ud} V_m/{\ud} z$ independent of both $R$ and $\theta$. In the following, we neglect beads with velocities less than $100\un{mm/s}$ since these are assumed to belong to the static bed which solidly rotates with the drum.

\begin{figure}
\centering
\includegraphics[ width=0.98\columnwidth]{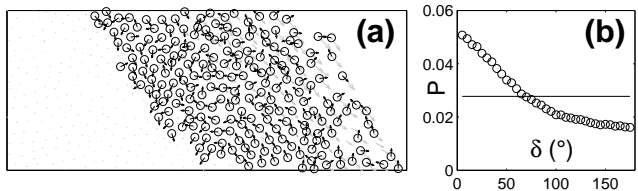}
\caption{(a) Typical instantaneous field of the orientation of the beads velocity fluctuations obtained for a quasi-2D flow with $\Omega=8\un{rpm}$. The orginal instantaneous velocity field is represented in gray. (b) Distribution function $P(\delta)$ of the angle $\delta$ between the orientation of the velocity fluctuations of two beads in contact. The straight line corresponds to a uniform fonction.} 
\label{fig2}
\end{figure}

{\em Velocity fluctuations in  quasi-2D and pure 2D flows:} In order to investigate possible collective effects, we study first the spatial correlations of the instantaneous velocity field for both geometry (A) and (B). Calling ${\bf V}(x,z,t)$ the velocity of a bead of the frame $t$ located at the $(x,z)$ coordinate in $({\bf e}_x,{\bf e}_z)$, where the unit vector ${\bf e}_x$ (resp. ${\bf e}_z$) is parallel (reps. perpendicular) to the mean flow, the bead fluctuation velocity is defined as: $\tilde{{\bf V}}(x,z,t)= {\bf V}(x,z,t)- V_m(z){\bf e}_x$. Since the orientation of the fluctuation $\tilde{\bf n}=\tilde{\bf  V}/\tilde{V}$ is not correlated to the depth $z$ (at variance with the fluctuation amplitude $\tilde{V}$), we focussed on the instantaneous field made up of these orientations.

\begin{figure}
\centering
\includegraphics[ width=0.98\columnwidth]{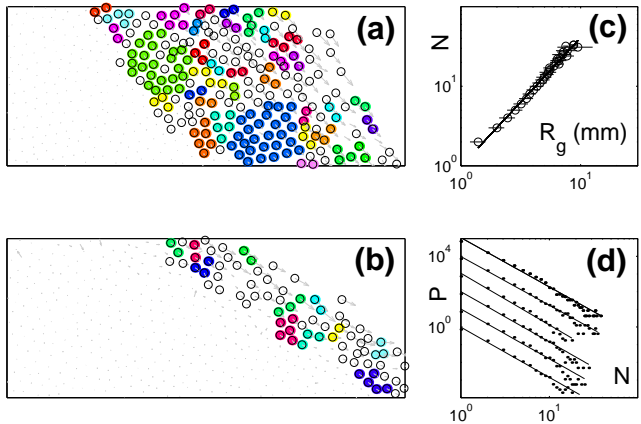}
\caption{Clusters of beads with correlated velocity fuctuation orientation in quasi 2D flows (geometry (A)). (a) and (b): Typical frames of the clusters respectively for $\Omega=8\un{rpm}$ and $\Omega=1\un{rpm}$ (see text for details en the cluster definition). The position of the beads belonging to the flowing layer are represented with a black circle. The gray arrows represent the raw velocity field. Beads belonging to the same cluster are drawn in the same color (c): cluster size $N$ plotted against their radius of gyration $R_g$. The errorbar indicated the standard deviation on $R_g$ of clusters with a given $N$. A power law behaviour (straight line) indicates fractal structures of dimension $d_f^{(A)}=1.6\pm 0.1$. (d): Probability density function of the size (in number of beads N) of these clusters. Data have been shifted for clarity. From bottom to top, $\Omega=1\un{rpm}$, $\Omega=2\un{rpm}$, $\Omega=3\un{rpm}$, $\Omega=4\un{rpm}$, $\Omega=6\un{rpm}$ and $\Omega=8\un{rpm}$. For these different $\Omega$, {\em i.e.} for different $\theta$ and $R$, the distribution decreases as a power-law (straight line) with the same exponent $\alpha^{(A)}=2.9\pm 0.1$.} 
\label{fig3}
\end{figure}

One of these instantaneous fields is represented on Fig.\ref{fig2}a and reveals aggregates of strongly correlated beads. 
These correlations can be quantified by looking at the distribution of the angle $\delta$ between
 the orientation $\tilde{\bf n}_i$ and $\tilde{\bf n}_j$ of two beads $i$ and $j$ in contact~\cite{defContact} (Fig.~\ref{fig2}b). This distribution reveals that the velocity 
fluctuations of two beads in contact tend to have correlated orientations. To isolate the correlated aggregates, we have used the following criteria: Two beads $i$ and $j$ belong to the 
same cluster whenever (1) they are in contact and (2) the angle between $\tilde{\bf n}_i$ and $\tilde{\bf n}_j$ is smaller than a given value arbitrary chosen equal to $60\,^\circ$. 
This value corresponds to the point where the experimental orientation distribution crosses the value of the uniform distribution (see Fig.~\ref{fig2}b). Values ranging from $30^\circ$ up to $90\,^\circ$ have also been tried and do not modify the following conclusions. The Figs.~\ref{fig3}a,b show typical frames of the individual clusters for geometry (A) respectively for $\Omega=8\un{rpm}$ and $\Omega=1\un{rpm}$. Note that the size of the largest ones are of order of the flowing layer thickness $R$. We also plot the number of beads $N$ in a cluster against the cluster radius of gyration $R_g$ (Fig.~\ref{fig3}c). The power law scaling observed in both geometries (A) and (B) is indicative of a fractal structure with an apparent fractal dimension equal respectively to $d_{f}^{(A)}=1.6\pm0.1$ and $d_{f}^{(B)}=1.9\pm0.1$ for quasi 2D and in pure 2D flows. This quantifies the observation that the clusters are more open and stringy in the quasi
2D geometry. More interestingly, the distribution of the number of beads $N$ in a cluster is a
power-law $N^{-\alpha}$ in both geometries (A) and (B), with an exponent equal respectively to
 $\alpha^{(A)}=2.9\pm 0.1$ and $\alpha^{(B)}=2. \pm 0.1$ for quasi-2D and 2D flows 
(see Fig.~\ref{fig3}d for the quasi-2D case). 
In both cases, the exponent is independent of $\Omega$, {\em i.e.} independent of both $R$ and $\theta$. This indicates that no typical correlation length scale can be defined to characterize the spatial correlations of the velocity field. 

\begin{figure}
\centering
\includegraphics[ width=0.98\columnwidth]{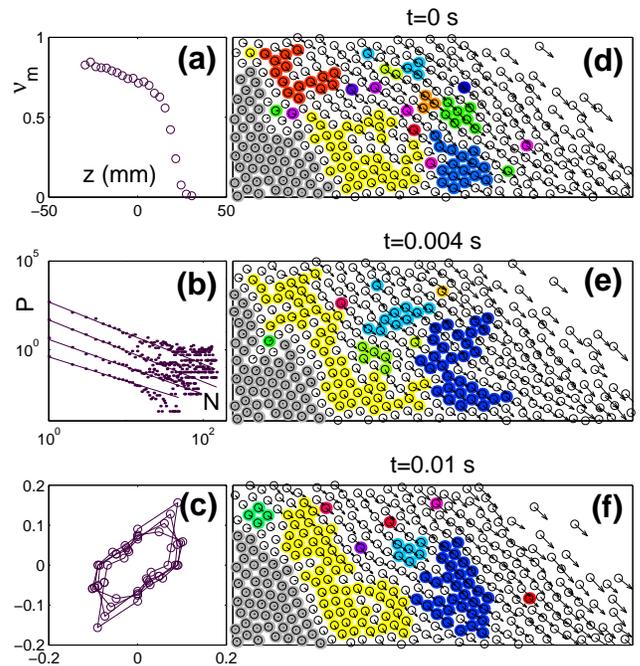}
\caption{Presence of rigid clusters in the flowing layer of pure 2D flows (geometry (B)). (a) Averaged volume fraction profile $\nu_m(z)$. (b) Probability density function of the size (in number of beads $N$) of the "rigid" clusters (see text for details). Data have been shifted for clarity. From bottom to top, $\Omega=2\un{rpm}$, $\Omega=7\un{rpm}$, $\Omega=10\un{rpm}$ and $\Omega=13\un{rpm}$. They are power law distributed with an exponent close to $\alpha_\nu^{(B)} \simeq 1.5$ weakly dependent of $\Omega$. (c) Polar diagram of the probability density of cluster direction, showing a preferential orientation parallel to the flow. Frames (d) (e) and (f) show a typical sequence of these clusters. Beads belonging to the static phase appear in gray. Beads belonging to the same cluster are drawn in the same color. The clusters are tracked on the successive frames.} 
\label{fig4}
\end{figure}

{\em Volume fraction fluctuations in pure 2D flows:} In order to find out the physical origin of 
these correlated clusters, measurements on the local volume fraction fluctuations were performed on 
the 2D stacking (geometry (B)). A Voronoi tesselation can then be used to define the local volume 
fraction $\nu$ associated with each bead. Let us note that this technique assumes that all beads position are known and consequently cannot be applied for the quasi-2D packing of geometry (A). 
The relative fluctuations of $\nu$ are very small, a few percent (see Fig.~\ref{fig4}a), but sufficient to alter
significantly the beads behaviour. According to Reynolds' dilatancy concept~\cite{Reynolds85}, an assembly of rigid particles cannot deform whenever its volume fraction is higher than a given threshold $\nu_c$. For pure 2D packing of monodisperse beads, $\nu_c=\pi/4$. Clusters of beads in contact with $\nu \geq  \nu_c$ have been isolated in the flowing layer (see fig.~\ref{fig4}d,e,f). Almost 50\% of beads of the flowing layer belong to one of these solid clusters. As was found for 
the clusters defined using velocity correlation, the size of largest ones is of order of $R$. The size distribution is again a power law with 
an exponent $\alpha_\nu^{(B)}\simeq 1.5$, weakly dependent of $\Omega$  (see Fig.~\ref{fig4}b). This exponent is slightly smaller than the one obtained from by analysis of velocity fluctuation. It should be noted that although there is some overlap between both types of clusters they are not identical. The fractal dimension of these `dense' clusters is found to be $d_\nu=1.7\pm0.1$. 
Their mean aspect ratio, calculated as the ratio of the two principal axis of their inertia tensor, does not depend on their size and is found to be close to $2$. 
They preferentially orient in the flow direction (see Fig.~\ref{fig4}c). Their life time, defined as the duration during which their size is larger than the half of their maximal size, is of order of $\sqrt{d/g}\simeq 10^{-2}\un{s}$, where $g$ is the gravity. 

{\em Discussion:} These solid clusters can be compared to the collective objects postulated in different non local models for granular flows. Mills et al.~\cite{Mills99} postulate the existence of 1D transient solid chains in the flowing layer, well separated from each other. In their approach, that describes granular flow down a rough inclined plane, a grain colliding with one of these chains transmits its momentum throughout the whole chain. On the other hand, for surface flows, Rajchenbach~\cite{Rajchenbach00} suggests that each shock impact is imparted to the whole substrate beneath and thus that the momentum is `dissipated' into the whole packing. 
This description has been recently
 extended by Andreotti and Douady~\cite{Andreotti01} who account for the possible trapping of flowing grains in the bumps of the static bed. The localization of the flow within a layer of finite thickness and the linear velocity profile are well reproduced in their approach. However their model does not account for the experimental dependence of the velocity gradient in the thickness $R$ and the local angle $\theta$. The collective objects that we observe are neither chain-like nor the whole packing, but rather widely distributed fractal clusters involving $50\%$ of the beads of the flowing layer. No typical length scale can be defined. This last point may be responsible for the selection of a velocity gradient independent of both $R$ and $\theta$, {\em i.e.} independent of 
the shear stress.

The observation of the successive clusterised frames shows that the largest clusters are 
emitted by the static phase and die either by fragmenting into smaller ones or by sticking back to the static phase. Differences between granular flows down a rough inclined plane and granular surface flows can then be rationalized as follows: clustering results from the competition between inelastic multiple collisions, that tends to aggregate grains together~\cite{Goldhirsch98,Falcon99}, and shear, that erodes clusters. For flows down an inclined plane, these two effects lead to clusters of a typical size, whereas in surface flows, the static bed plays the role of a cluster reservoir: its erosion by the flowing grains can generate very large size clusters that then split and cascade into smaller and smaller ones (this mechanism might explain why we observe 
power-law distributions). In this view, the cluster size distribution, and a fortiori the velocity gradient, depend crucially on the boundary conditions.

The absence of a characteristic correlation length to describe the locally `jammed' clusters
of beads is quite interesting. First, this contrasts with an assumption made in most local
and non local models~\cite{Aranson01,Mills99,Rajchenbach00,Andreotti01}, where the transition
between the `solid' phase and the `liquid' phase is supposed to occur over a well defined length
scale. The existence of multi-scale rigid clusters indicates that the flowing phase is
actually {\it critical}, and suggests the proximity of a continuous `jamming' transition, of the
type proposed in ~\cite{Liu98} (see also ~\cite{Halsey01} for a related discussion). In this respect, it is interesting to note that similar power-law distributed clusters of strongly correlated motion have recently been observed in a colloid close to the glass transition~\cite{Weeks00}. 

\begin{acknowledgments}
We thank B. Dubrulle, O. Pouliquen and D. Salin for
fruitful discussions, and C. Gasquet, V. Padilla and P. Meininger for technical
assistance.
\end{acknowledgments}


\begin{thebibliography}{99}
\bibitem{defContact} For pure 2D flows, the definition of a criterion for bead contact does 
not pose difficulties~: two beads are in contact as soon as the distance separating their center of mass is smaller than $1.1d$. The factor $1.1$ accounts for the precision in the experimental location of the beads. For quasi 2D flows, it is more complicated since the distances between beads and the side wall of the tumbler are unknown: However, investigations on the distribution of the experimentally measured distances between a beads and its closest neighbours convince us that most beads actually seen by the camera are in contact this the tumbler side wall. We thus have kept the same thershold for quasi-2D flows.
\end{thebibliography}
\end{document}